\documentclass[aps,pra,twocolumn,groupedaddress,floatfix,showpacs]{revtex4-2}
\usepackage{graphicx} 
\usepackage{amsmath, braket}
\usepackage{amssymb}
\usepackage{color}

\begin{document}

\title{Chirality-Induced Spin Currents in a Fermi Gas}
\author{Camen A. Royse and J. E. Thomas}

\affiliation{Department of Physics, North Carolina State University, Raleigh, NC 27695, USA}

\date{\today}

\begin{abstract}
We observe and model spin currents arising from chirality and effective spin-exchange interactions in a weakly interacting $^6$Li Fermi gas. 
Chirality is introduced by a static displacement between the center of the trapped atoms and the center of an applied magnetic bowl, which produces left- or right-handed spatially varying spin rotation. Spin current is directly observed via oscillations in the centers of mass of the spin-up and spin-down components, which appear to bounce off of or pass through one another, depending on the degree of handedness and s-wave scattering length. We show that this behavior obeys a driven oscillator equation with an effective spin-dependent driving force. Our measurements demonstrate chirality-induced spin selectivity via the direction of the current flow, extending CISS phenomena to Fermi gases. 
\end{abstract}

\maketitle

Chiral (twisted) materials are of great interest for generating and controlling spin-polarized electron currents~\cite{RayCISS2} to enable ``spintronics'' technologies. In such technologies, information is transmitted via spin currents instead of electron currents, greatly reducing Joule heating, which limits the scalability of computers. Chiral molecules~\cite{ShangChiralMolec} can act as passive spin filters, efficiently transporting electrons for one polarization while impeding the transmission of the other. Large chiral-induced spin selectivity (CISS) has been observed in the asymmetric transmission of polarized electrons through thin films of chiral molecules, including DNA~\cite{SpinTransportDNA,CISSDNA2}. While spin-orbit coupling  is known to play a role in this effect, no unifying theory of the role of chirality has been able to quantitatively reproduce measurements. Spin transport in chiral materials is now widely studied~\cite{BloomCISS,Chiralviewpoint,ChiralSolid-1,MoharanaCISS1} for manipulating spin-selective currents. Ultracold Fermi gases provide experimental simulators for spintronics concepts~\cite{DuineHighlight2008}, which may offer new insights into chiral control methods. 

\begin{figure}[htb]
\includegraphics[width=3.15in]{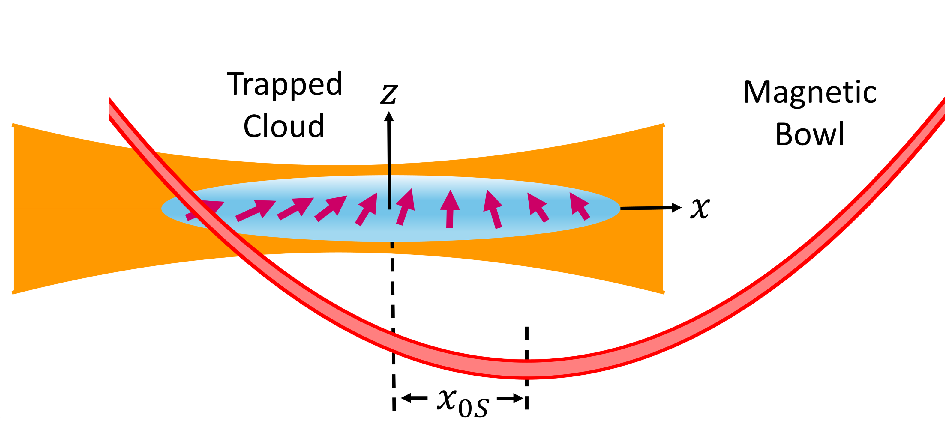}
    \caption{Controlling spin twist in an atomic Fermi gas. A cigar-shaped cloud of $^6$Li (light blue), prepared in a superposition of spin-up and spin-down states, is confined by a focused CO$_2$ laser beam (orange) and a magnetic bowl (red). $x$-dependent spin twist about the chiral $z$-axis  is controlled by the offset $x_{os}$ between the centers of the spin-dependent magnetic bowl potential and the trapped cloud (blue) at $x\equiv 0$. }
    \label{fig:trap}
\end{figure}

We directly observe chirality-induced spin currents in a trapped, weakly interacting Fermi gas of $^6$Li atoms, confined in the combined potential of a cigar-shaped CO$_2$ laser trap and a magnetic bowl, Fig.~\ref{fig:trap}. In contrast to chiral materials, where twist is structural, a chiral spin texture (i.e., a spin spiral/twist) is created in the cloud by simply displacing the center of the laser trap from the center of the magnetic bowl, which causes the coherently prepared spin medium to evolve into a twisted state. We show that the spin centers of mass obey a 1D driven oscillator equation, where an effective spin-dependent driving force arises from s-wave scattering in the twisted medium and is tunable from aperiodic to impulsive. 


The $^6$Li atoms are initially prepared in a $z$-polarized spin state $\ket{\uparrow}$. A $0.5$ ms $\pi/2$ rf pulse then creates a pseudospin state~\cite{BW}, which we define to be $x$-polarized, $(\ket{\uparrow}+ \ket{\downarrow})/\sqrt{2}$, where $\ket{\uparrow}\,\ket{\downarrow}$ denote the two lowest hyperfine states, in order of increasing energy.   A bias magnetic field $B_z$ (orthogonal to the trap axis $x$) both defines the quantization axis and controls spin-spin interactions by tuning the $\ket{\uparrow}$-$\ket{\downarrow}$  $s$-wave scattering length $a_s$. Key to the experiments is the curvature of $B_z(x)$, which creates a spin-dependent harmonic potential (i.e., a magnetic bowl) that confines the cloud along the $x$-axis, with negligible effect in the narrow transverse directions $y,z$.  The CO$_2$ laser potential provides radial confinement and has a relatively short Rayleigh length, $\simeq 0.75$ mm,  producing a spin-independent axial force that is comparable to that of the magnetic bowl.  The center of the magnetic bowl is determined by fixed electromagnets, while the  CO$_2$ laser focal point is easily translated to shift the center of the trapped atoms relative to that of the magnetic bowl.

The net optical and magnetic axial potential is
\begin{equation}
V_{\uparrow,\downarrow}(x)=\frac{m\,\omega_0^2}{2}x^2\pm\frac{m\omega_0\delta\omega_{\uparrow\downarrow}}{2}(x-x_{os})^2\,,
\label{eq:potentials}
\end{equation}
where $\omega_0\equiv 2\pi\nu_x\simeq 2\pi\times 25$ Hz is the average harmonic oscillation frequency and $\delta\omega_{\uparrow\downarrow}\simeq 2\pi\times 14$ mHz  is the difference in the harmonic oscillation frequencies for the two states~\cite{Deltaomega}.
Note that the small value of $\delta\omega_{\uparrow\downarrow}/\omega_0$ is due to the fact that the two states differ only in nuclear spin projection. The average  potential is spin-independent and centered at $x\equiv 0$. Chirality is controlled by $x_{os}$,  the offset between the measured centers of the magnetic bowl and the trapped cloud.

In equilibrium, $F_{\uparrow,\downarrow}=-\partial_xV_{\uparrow,\downarrow}(x)=0$, the displacement between the centers of the density profiles $n_\uparrow$ and $n_\downarrow$, to first order in $\delta\omega_{\uparrow\downarrow}/\omega_0$, is
\begin{equation}
 x_{\uparrow} - x_{\downarrow}=\frac{2\delta\omega_{\uparrow\downarrow}}{\omega_0}\,x_{os}\,.
 \end{equation}
 For our experiments, where $x_{os}\simeq\sigma_x\simeq 200\,\mu$m, we find $x_{\uparrow} - x_{\downarrow}\simeq 0.2\,\mu$m,  which is negligible compared to the cloud widths. We observe no difference in the positions of pure $\ket{\uparrow}$  or $\ket{\downarrow}$ states in the trap.  Hence, the difference between the mechanical forces does not play a role in the evolution of the system. However, the spatially varying frequency shift $(V_\uparrow-V_\downarrow)/\hbar$ produces a spatially varying rotation rate of the pseudospins about the chiral $z$-axis
 \begin{equation}
{\mathbf \Omega }=\hat{\mathbf z}\,\Omega_0 \frac{(x-x_{os})^2}{\sigma_x^2}\,,
\label{eq:Omegaz}
\end{equation}
 in a frame rotating at the hyperfine resonance frequency, creating a twisted spin state as the system evolves~\cite{SpinSpiral}. Here $\Omega_0\equiv m\omega_0\delta\omega_{\uparrow\downarrow}\sigma_x^2/\hbar=52.6\,s^{-1}$ at 511 G for $\sigma_x=198\,\mu$m~\cite{SupportOnline}. $\Omega_0$ is determined by the known magnetic field tuning rate for each state and the measured oscillation frequency of atoms in the magnetic bowl $\omega_{\rm mag}$~\cite{Deltaomega}.
   
 Just after the $\pi/2$ pulse, all of the spins are  parallel~\cite{BW} and hence non-interacting in a Fermi gas restricted to $s$-wave scattering. The initial evolution of the pseudo-spins is then governed by the $x$-dependent rotation rate  ${\mathbf \Omega}$, causing the spin vectors to point in different directions at a later time, enabling scattering interactions between the atoms.  In the weakly interacting regime, the atoms remain in simple harmonic motion. However, forward $s$-wave scattering between atoms with nonparallel spins causes each spin vector to rotate about the total spin vector~\cite{LaloeISRE,Fuchs,Piechon,DuSpinSeg2}, analogous to a spin-exchange interaction, which produces a measurable spin $z$ component.

\begin{figure}[htb]
\includegraphics[width=2.5in]{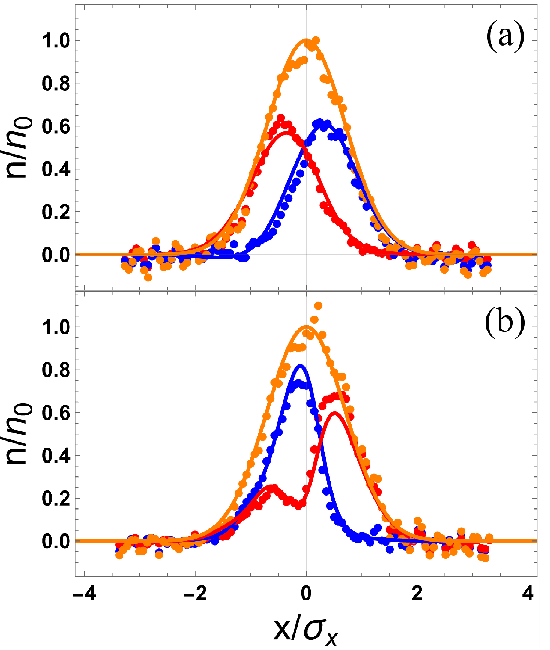}
    \caption{Spatial profiles for spin-up and spin-down states.  Spin-up (blue),  spin-down (red). The total density $n_\uparrow+n_\downarrow$  (orange) is conserved.  Chirality is controlled by the offset parameter $x_{os}$, see Fig.~\ref{fig:trap}. a)  $x_{os}=1.0\,\sigma_x$ at $t=21$ ms with an s-wave scattering length $a_s=-53.1\,a_0$ G; b) $x_{os}=-0.9\,\sigma_x$ at $t=37$ ms for $a_s=-213.1\,a_0$. Predictions (solid curves).}
    \label{fig:Spatial511437}
\end{figure}

 We investigate the interplay between the interaction strength, controlled by the $s$-wave scattering length $a_s$, and the chirality, controlled by the offset $x_{os}$.  Asymmetric spatial profiles are observed for nonzero static displacements $x_{os}$, as shown in Fig.~\ref{fig:Spatial511437}. The experiments are performed in the nondegenerate regime, where the reduced temperature $0.5<T/T_F<0.7$. The profiles of both spin states are identical immediately after the rf pulse and nominally gaussian, centered at $x=0$ with a $1/e$ width $\sigma_x\simeq 200\,\mu$m~\cite{SupportOnline}. 
 
Fig.~\ref{fig:Spatial511437}(a) shows density profiles for $x_{os}=1.0\,\sigma_x$ at $B = 511$ G, where the s-wave scattering length $a_s=-53.1\,a_0$, the peak total density $\bar{n}_0=0.90/\mu m^3$ (see below), and the trap frequency $\nu_x=25.9$ Hz.  At $t=21$ ms after the $\pi/2$ rf pulse (about half of the trap period), the spin-up profile (blue) is shifted to the right, while spin-down profile (red) is shifted left~\cite{Reversal}. Fig.~\ref{fig:Spatial511437}(b) shows the results obtained for negative $x_{os}=-0.9\,\sigma_x$ with  at  $B=437$ G, where $a_s=-213.1\,a_0$, $\bar{n}_0=0.52/\mu m^3$, and $\nu_x=24.3$ Hz. At $t=37$ ms, the profile of the spin-up (blue) is shifted to the left, while that of the spin-down component (red) is distorted on the left and shifted right. In both cases, the total spin density $n(x)=n_\uparrow(x,t)+n_\downarrow(x,t)$ is time-independent and remains a gaussian centered at $x=0$ (orange), demonstrating spin current without total density (charge) current. Solid curves show the predictions of the model, discussed below.

\begin{figure*}
\begin{center}\
\hspace*{-0.25in}
\includegraphics[width=6.0in]{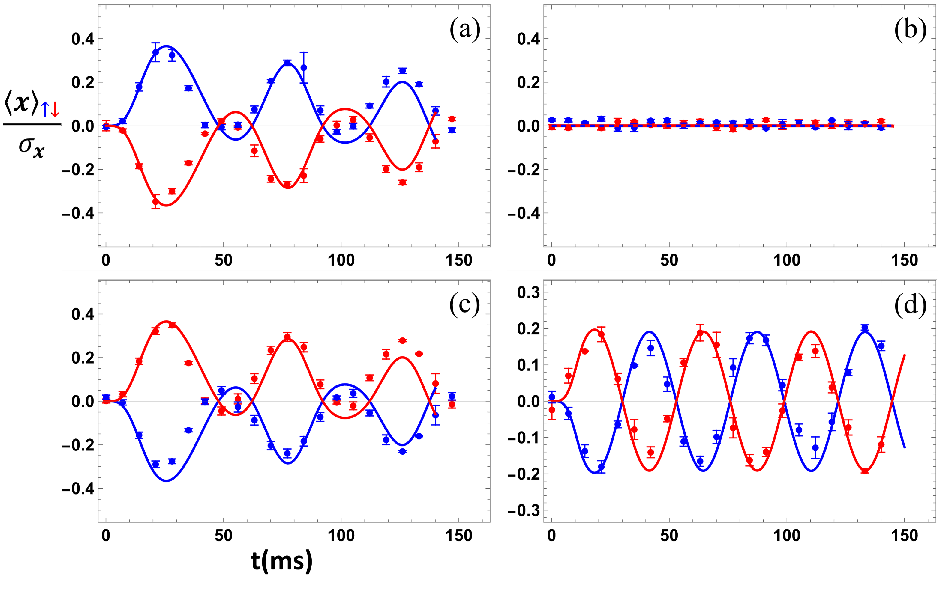}
\end{center}
\caption{Oscillating spin current versus chirality. $\langle x\rangle_{\uparrow\downarrow}$ denotes the center-of-mass of each spin state (blue and red)~\cite{ErrorBar}.  The  s-wave scattering length is fixed at $a_s=-53.1a_0$. Chirality is controlled by the offset $x_{os}$, see Fig.~\ref{fig:trap}. (a) $x_{os}= 0.8\,\sigma_x$; (b) $x_{os}= 0.0$; (c) $x_{os}=- 0.8\,\sigma_x$; (d) $x_{os}=-3.0\,\sigma_x$, where $\sigma_x\simeq 200\,\mu$m.   Time is measured from the end of the $\pi/2$ RF pulse. The peak total density (a-c) is $\bar{n}_0=0.9/\mu m^3$, $\nu_x=25.9$ Hz; (d) $\bar{n}_0=0.42/\mu m^3$, $\nu_x=21.8$ Hz. The spin components reverse roles with the sign of $x_{os}$. Solid curves show predictions.}
\label{fig:combined511P1-M3}
\end{figure*}

We quantify the effect of chirality on the motion by plotting the center-of-mass  $\langle x\rangle_{\uparrow,\downarrow}=\int dx\, x\, n_{\uparrow,\downarrow}(x,t)/N_{\uparrow,\downarrow}$ for each state versus time, for different offsets $x_{os}$.  Fig.~\ref{fig:combined511P1-M3} shows data for $x_{os}$ of $0.8\,\sigma_x,0,-0.8\,\sigma_x$, and $-3\,\sigma_x$ at 511 G, where $a_s=-53.1\,a_0$. Fig.~\ref{fig:combined511P1-M3}(b) locates the center of the magnetic bowl, $x_{os}=0$, where there is no separation in the centers-of-mass. For $x_{os}\neq 0$, the centers of mass exhibit  $180^{\,o}$ out-of-phase spin-dipole modes. The directions of motion of the spins are interchanged when the sign of $x_{os}$ is reversed, Fig.~\ref{fig:combined511P1-M3} (a,c). With $x_{os}=\pm 0.8\,\sigma_x$,  the spin-up and spin-down profiles separate, return to the center, and appear to bounce off of one another. In contrast, for $x_{os}=-3\,\sigma_x$, Fig.~\ref{fig:combined511P1-M3}(d), the  centers of mass oscillate completely through one another.  




For $x_{os}=0$, symmetrical spin-segregation has been observed in the spin density profiles~\cite{DuSpinSeg1,DuSpinSeg2,SaeedPRASpinECorrel,ThywissenDynamicalPhases}, where spin-up (down) atoms appear to move to the center (edges) of the trap,  reversing roles with the sign of the scattering length $a_s$. 
For the very small $a_s<10\,a_0$ used in previous experiments~\cite{DuSpinSeg1,DuSpinSeg2,SaeedPRASpinECorrel,ThywissenDynamicalPhases},  the evolution time of the spin density is long compared to the trap period. Then, the system can be modeled as an energy-space lattice~\cite{DuSpinSeg2,SaeedPRASpinECorrel} with effective long range interactions that arise from time-averaged s-wave scattering between atoms oscillating in the trap~\cite{LewensteinDynLongRange}. With $x_{os}=0$,  where $\langle\Omega\rangle\propto\langle x^2\rangle\propto E$  from eq.~\ref{eq:Omegaz}, the spin vectors are correlated with the energy, yielding spatially symmetric spin density profiles~\cite{DuSpinSeg2,SaeedPRASpinECorrel,KollerReySpinDep,WallEnergySpinLattice,ThywissenDynamicalPhases,JingjingCorrelation,RoyseLightLossPRL}.

In the present work, we observe tunable spin center of mass motion for  $x_{os}\neq 0$ on time scales fast compared to the trap period, the opposite limit, by employing larger scattering lengths $a_s>50\,a_0$ and spin rotation rates that can exceed the trap frequency. Here, the effective long range interaction picture is not valid and the spin density profiles are spatially asymmetric. 

We employ a 1D mean field treatment~\cite{SupportOnline} of the weakly interacting cloud, where momentum-changing collisions can be neglected for the measurement time scale.  The Heisenberg equations of motion are used to determine the evolution of the spin density components $(s_x,s_y,s_z)\equiv{\mathbf s}$  and the corresponding spin current densities $(J_x,J_y,J_z)\equiv{\mathbf J}$ in a one dimensional approximation, where the direction of  ${\mathbf J}$ denotes the direction of the spin carried by the current flowing along the x-axis.   For our experimental parameters, forces arising from spatial derivatives of the mean field potential are negligible, as is the difference of the magnetic forces. In this case, the evolution equation for the total density shows that $n(x,t)=n_\uparrow(x,t)+n_\downarrow(x,t)=n_0(x)$ is time-independent, as observed in the experiments. With these approximations, and assuming a local Maxwell-Boltzmann momentum distribution at temperature $T$, the spin density ${\mathbf s}(x,t)$ and current ${\mathbf J}(x,t)$ obey
\begin{equation}
\dot{\mathbf s}+\partial_x\,{\mathbf J}={\mathbf \Omega}\times{\mathbf s},
\label{eq:S}
\end{equation}
\begin{equation}
\dot{\mathbf J}+\frac{k_BT}{m}\partial_x\,{\mathbf s}+\omega_0^2 x\,{\mathbf s}={\mathbf \Omega}\times{\mathbf J}-\frac{8\pi \hbar a_s}{m}\,{\mathbf s}\times{\mathbf J}\,.
\label{eq:J}
\end{equation}
Here, ${\mathbf \Omega}$ (eq.~\ref{eq:Omegaz}) models the twist and $a_s\,{\mathbf s}\times{\mathbf J}$ governs the spin-selective mean field rotation rate of ${\mathbf J}$ about ${\mathbf s}$. 

Just after the $\pi/2$ radio-frequency pulse, the only non-vanishing initial condition  is $s_x(x,0)=\frac{1}{2}\,n_0(x)$, where $n_0(x)=\bar{n}_0\,\exp (-x^2/\sigma_x^2)$ and $k_BT/m=\omega_0^2\sigma_x^2/2$. The $s$-wave scattering lengths at each magnetic field are calculated from the online data of ref.~\cite{JochimPreciseFeshbach}. The one-dimensional approximation implies that the strength of the  ${\mathbf s}\times{\mathbf J}$ interaction term, which is proportional to the square of the density, should be averaged over the transverse dimensions. To fit the data, we find that the average peak 3D density $\bar{n}_0$ lies between $n_0/2$, as expected for a gaussian distribution, and $n_0$, the peak 3D density, calculated from the measured total atom number and cloud profile~\cite{SupportOnline}. Numerical solution determines the spin density profiles, $n_{\uparrow,\downarrow}(x,t)=\frac{1}{2}\,n_0(x)\pm s_z(x,t)$, which requires that $\langle x\rangle_\downarrow=-\langle x\rangle_\uparrow$, as predicted previously~\cite{LewensteinDynLongRange} and observed in our experiments.



\begin{figure}[htb]
\hspace*{-0.25in}\includegraphics[width=3.00in]{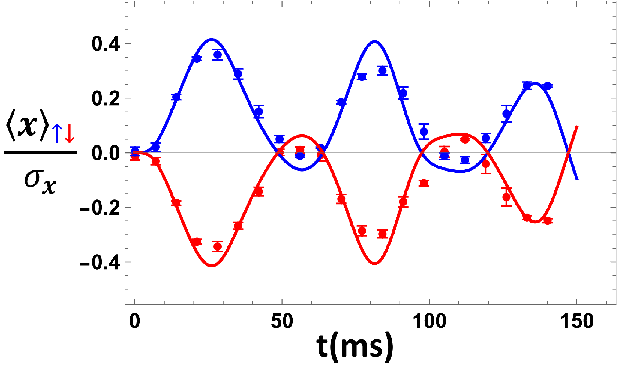}
    \caption{Oscillating spin current~\cite{ErrorBar} for an s-wave scattering length $a_s=-84.6\,a_0$ ($B=500$ G). The chirality parameter (see Fig.~\ref{fig:trap}) is $x_{os} = 1.15\,\sigma_x$ with $\sigma_x=197\,\mu$m, and $\bar{n}_0=0.99/\mu m^3$ with $\nu_x=22.4$ Hz.}
    \label{fig:500GOsc}
\end{figure}

\begin{figure}[htb]
\includegraphics[width=3.00in]{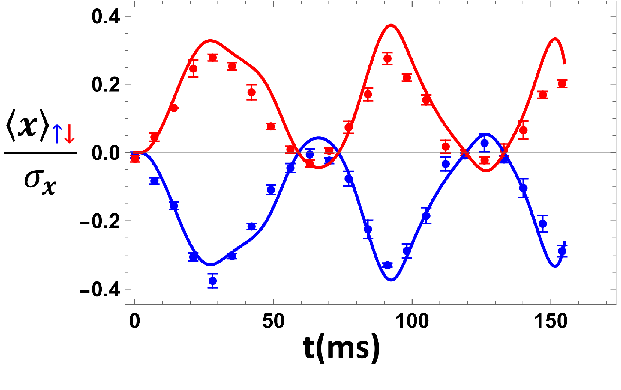}
    \caption{Oscillating spin current~\cite{ErrorBar} for an s-wave scattering length $a_s=-213.1\,a_0$ ($B=437$ G). The chirality parameter (see Fig.~\ref{fig:trap}) is $x_{os} =-0.8\,\sigma_x$ with $\sigma_x=195\,\mu$m, and $\bar{n}_0=0.52/\mu m^3$ with $\nu_x=24.3$ Hz.}
    \label{fig:437GOsc}
\end{figure}

The $z$-components of eqs.~\ref{eq:S}~and~\ref{eq:J} show that the centers of mass obey driven oscillator equations~\cite{SupportOnline}, 
\begin{equation}
\partial^2_t\langle x\rangle_\uparrow +\omega_0^2 \langle x\rangle_\uparrow=-\frac{8\pi \hbar a_s}{m N_\uparrow}\int\! dx\, ({\mathbf s}\times{\mathbf J})_z\,,
\label{eq:DrivenOsc}
\end{equation}
where the effective driving force on the right side has an opposite sign for the $\downarrow$ spin component. In eq.~\ref{eq:DrivenOsc}, $N_{\uparrow,\downarrow}=\bar{n}_0\sigma_x\sqrt{\pi}/2$ for a gaussian spatial profile. 

Fig.~\ref{fig:combined511P1-M3}(d) shows the fit of the model for the smallest scattering length $a_s=-53.1\,a_0$ (511 G), lowest density, and largest offset $x_{os}=-3\,\sigma_x$. In this case, the spatially varying rotation rate ${\mathbf \Omega}$ is nearly linear in $x$ and dominates the mean field rotation rate $\propto a_s$ in eq.~\ref{eq:J}. After the $\pi/2$ pulse,  atoms scatter from a transient spin spiral, see Fig.~S2~\cite{SupportOnline}. The $x$ and $y$ components of ${\mathbf s}$ and ${\mathbf J}$ form and vanish quickly compared to the trap period $1/\nu_x$,  due to the rapidly increasing spread in rotation angles for atoms at different positions. As spin exchange scattering vanishes for the resulting $z$-polarized spin density and current, the effective force on the right hand side of eq.~\ref{eq:DrivenOsc} takes the form of a pulse near $t=0$, see Fig.~S1~\cite{SupportOnline}. This produces an impulsive oscillator response with a period  near $1/\nu_x=45.8$ ms.

Fig.~\ref{fig:combined511P1-M3}(a,c), Fig.~\ref{fig:500GOsc}, and Fig.~\ref{fig:437GOsc} show the  spin current obtained when the mean field rotation rate is larger than  $\nu_x$ and comparable to or larger than the spread in ${\mathbf \Omega}$ across the trapped cloud. For these data, the model shows that the time-dependent effective force continuously modulates the oscillation amplitude and has a component that shifts the centers of oscillation, see Fig.~S1~\cite{SupportOnline}, causing the apparent bouncing.  With increased s-wave scattering lengths, where ${\mathbf J}$ and ${\mathbf s}$ are more strongly coupled, the oscillation is distorted from pure sinusoidal.


In summary,  our measurements demonstrate a general mechanism for manipulating the flow of spin current via asymmetric scattering in a chiral medium.  The vanishing of s-wave scattering for parallel spins, according to the Pauli principle, makes the twisted Fermi gas cloud spin orienting. In chiral molecules, the CISS effect remains puzzling because the electron velocity through the crystal electric fields can be spin dependent even without atomic spin-orbit coupling.  For the Fermi gas, there is no spin orbit coupling and the atom motion is unaffected, as the difference between the spin-dependent mechanical forces is negligible. Nevertheless, spatially asymmetric (chiral) spin waves appear as oscillations in the centers of mass of the spin components, driven by an effective force, which includes the back reaction of the spin current on the spin density.  Our work motivates new Fermi gas quantum simulators with tailored chirality, which could include synthetic spin-orbit coupling~\cite{SpielmanSO}.

Primary support for this research is provided by the Air Force Office of Scientific Research (FA9550-22-1-0329). Additional support is provided by the National Science Foundation (PHY-2307107). We are pleased to acknowledge Prof.~Dali Sun for encouragement in this work.

%

\newpage
\widetext
\setcounter{figure}{0}
\setcounter{equation}{0}
\renewcommand{\thefigure}{S\arabic{figure}}
\renewcommand{\theequation}{S\arabic{equation}}

\appendix

\section{Supplemental Material}
In this supplemental material, we describe a simplified one dimensional model of a twisted weakly interacting Fermi gas and show that the centers of mass for each spin component obey a driven harmonic oscillator equation with an effective spin-dependent driving force.  We tabulate the parameters used to model the data and display the effective forces.  

\subsection{Mean-Field Model of a Twisted Fermi Gas}
\label{sec:MFmodel}

Mean field models appropriate for weakly interacting Bose and Fermi gases have been developed by several groups~\cite{Williams,Levitov,DuSpinSeg1,LewensteinDynLongRange,LaloeSpinReph,LaloeISRE}. Our interest is in the nearly collisionless regime, where the density and s-wave scattering length are sufficiently small that momentum-changing collisions and damping are negligible for the observation period~\cite{DuSpinSeg2,MuellerWeaklyInt,Piechon,SaeedPRASpinECorrel}. Here, we summarize a simplified 1D model of a  weakly interacting Fermi gas tailored to our experiments, which employ a harmonically trapped, coherently prepared sample.  In our experiments, a spatially varying spin rotation rate is introduced by a static displacement between the center of a spin-dependent magnetic potential and the center of a spin-independent average potential, where the atoms are trapped.  S-wave scattering in the resulting spin twisted medium produces spontaneously oscillating spin waves and corresponding spin-currents, similar to those predicted for a spin spiral state~\cite{LewensteinDynLongRange}. We directly observe the oscillation in the motion of the centers of mass of each spin component, which we show obey a driven oscillator equation with a spin-dependent effective driving force.

The single particle Hamiltonian for each hyperfine spin state $\sigma\equiv\,\uparrow,\downarrow$, is
\begin{equation}
H_{0\sigma}({\mathbf x})=-\frac{\hbar^2}{2m}\nabla^2+V_\sigma({\mathbf x})+E_\sigma\,,
\label{eq:1.1S}
\end{equation}
where $E_\sigma$ denotes the hyperfine-Zeeman energy of each state, for a bias magnetic field $B=B_0$ at the center of the trapped cloud. As discussed in the main text, the net spin-dependent trapping $V_{\uparrow\downarrow}$ is 
\begin{equation}
V_{\uparrow,\downarrow}({\mathbf x})=\frac{m\,\omega_\perp^2}{2}(y^2+z^2 ) +\frac{m\,\omega_0^2}{2}x^2\pm\frac{m\omega_0\delta\omega_{\uparrow\downarrow}}{2}(x-x_{os})^2\,.
\label{eq:potentialsS}
\end{equation}
Here, the $x^2$ term is the average optical and magnetic axial potential, which is spin-independent and centered at $x_{mech}\equiv 0$. We define  $\omega_0=2\pi\nu_x$, where $\nu_x\simeq 25$ Hz is the typical axial oscillation frequency. The magnetic bowl has a negligible effect on the transverse potential, where $\omega_\perp=2\pi\nu_r$ with $\nu_r\simeq 625$ Hz the transverse oscillation frequency.  The spin-dependent potential arises only from the magnetic bowl, which is centered at $x_{mag}$. We define $x_{os}=x_{mag}-x_{mech}$ as the distance between the measured centers of the magnetic bowl and the trapped cloud. The spin dependent potential arises from the difference between the magnetic tuning rates for the two states and is given in terms of the difference in the net harmonic oscillation frequencies, $\delta\omega_{\uparrow,\downarrow}\simeq 2\pi\times 14$ mHz~\cite{SaeedPRASpinECorrel,Deltaomega}. The spin-dependent contribution to the tightly confining transverse potential is negligible.    

We begin with the Schr\"{o}dinger picture Hamiltonian, $\hat{H}=\hat{H}_0+\hat{H}_{int}$, with
\begin{equation}
\hat{H}_0=\sum_\sigma \int d^3{\mathbf x}'\,\hat{\psi}^\dagger_\sigma({\mathbf x}')H_{0\sigma}({\mathbf x}')\hat{\psi}_\sigma({\mathbf x}')\,
\label{eq:1.2S}
\end{equation}
where the field operators $\hat{\psi}_\sigma({\mathbf x}')$ obey the standard anticommutation relations. 
The interaction Hamiltonian for short-range (contact) s-wave scattering is
\begin{equation}
\hat{H}_{int}=g_{12} \int d^3{\mathbf x}'\,\hat{\psi}^\dagger_\uparrow({\mathbf x}')\hat{\psi}^\dagger_\downarrow({\mathbf x}')\hat{\psi}_\downarrow({\mathbf x}')\hat{\psi}_\uparrow({\mathbf x}')\,,
\label{eq:1.4S}
\end{equation}
where 
\begin{equation}
g_{12}\equiv\frac{4\pi\hbar^2 a_s}{m}
\label{eq:1.6S}
\end{equation}
with $a_s=a_{12}$ the s-wave scattering length.

The evolution equations for the spin density and spin current are found using the Wigner phase space operators,
\begin{equation}
\hat{W}_{\alpha ,\alpha'}({\mathbf x},{\mathbf p})=\int\frac{d^3 \boldsymbol\epsilon}{(2\pi\hbar)^3}\,e^{i\frac{\boldsymbol\epsilon\cdot{\mathbf p}}{\hbar}}\,\hat{\psi}^\dagger_\alpha\left({\mathbf x}+\frac{\boldsymbol\epsilon}{2}\right)\hat{\psi}_{\alpha'}\left({\mathbf x}-\frac{\boldsymbol\epsilon}{2}\right)\,.
\label{eq:1.7S}
\end{equation}
The density matrix operator is then
\begin{equation}
\hat{\rho}_{\alpha ,\alpha'}({\mathbf x})=\int d^3{\mathbf p}\,\hat{W}_{\alpha ,\alpha'}({\mathbf x},{\mathbf p})=\hat{\psi}^\dagger_\alpha({\mathbf x})\hat{\psi}_{\alpha'}({\mathbf x})
\label{eq:1.8S}
\end{equation}
with the corresponding current density
\begin{equation}
{\mathbf J}_{\alpha ,\alpha'}({\mathbf x})=\int d^3{\mathbf p}\,\frac{\mathbf p}{m}\,\hat{W}_{\alpha ,\alpha'}({\mathbf x},{\mathbf p})\,.
\label{eq:1.9S}
\end{equation}

In a mean field approximation, we write the thermal averaged interaction Hamiltonian using Wick's theorem and allow first order fluctuations in each thermal average factor, 
\begin{eqnarray}
\hat{H}_{int}&=&g_{12} \int d^3{\mathbf x}'\,\left\{\bar{n}_\uparrow({\mathbf x}')\hat{\psi}^\dagger_\downarrow({\mathbf x}')\hat{\psi}_\downarrow({\mathbf x}')+\bar{n}_\downarrow({\mathbf x}')\hat{\psi}^\dagger_\uparrow({\mathbf x}')\hat{\psi}_\uparrow({\mathbf x}')\right.\nonumber \\
& &\hspace*{0.5in}\left. -\bar{\rho}_{\uparrow\downarrow}({\mathbf x}')\hat{\psi}^\dagger_\downarrow({\mathbf x}')\hat{\psi}_\uparrow({\mathbf x}')-\bar{\rho}_{\downarrow\uparrow}({\mathbf x}')\hat{\psi}^\dagger_\uparrow({\mathbf x}')\hat{\psi}_\downarrow({\mathbf x}')\right\}\,,
\label{eq:2.6S}
\end{eqnarray}
where the bar denotes a thermal average, $\bar{n}_\sigma({\mathbf x}')=\langle\hat{\psi}^\dagger_\sigma({\mathbf x}')\hat{\psi}_\sigma({\mathbf x}')\rangle$, $\bar{\rho}_{\sigma\sigma'}({\mathbf x}')=\langle\hat{\psi}^\dagger_\sigma({\mathbf x}')\hat{\psi}_{\sigma'}({\mathbf x}')\rangle$. Later, we restore the time-dependence to the thermal average factors, to be consistent with the thermal averaged Heisenberg picture operator evolution equations.

\subsubsection{Heisenberg Equations}

Now we can find the Heisenberg equations for the Wigner phase space operators $\hat{W}_{\alpha ,\alpha'}({\mathbf x},{\mathbf p},t)=U^\dagger(t)\,\hat{W}_{\alpha ,\alpha'}({\mathbf x},{\mathbf p})\,U(t)$. Suppressing the time $t$ and evaluating the equal time commutators in the Schr\"{o}dinger picture, we require
 \begin{equation}
\partial_t\hat{W}_{\alpha ,\alpha'}({\mathbf x},{\mathbf p})=\frac{i}{\hbar}[\hat{H},\hat{W}_{\alpha ,\alpha'}({\mathbf x},{\mathbf p})]\,.
\label{eq:4.2S}
\end{equation}
We define 
\begin{equation}
[\hat{H},\hat{\psi}^\dagger_\alpha({\mathbf x}_+)\hat{\psi}_{\alpha'}({\mathbf x}_-)]=\hat{\psi}^\dagger_\alpha({\mathbf x}_+)[\hat{H},\hat{\psi}_{\alpha'}({\mathbf x}_-)]-[\hat{H},\hat{\psi}_{\alpha}({\mathbf x}_+)]^\dagger\hat{\psi}_{\alpha'}({\mathbf x}_-)\equiv C^{\,0}_{\alpha\alpha'}+C^{\,int}_{\alpha\alpha'}
\label{eq:4.4S}
\end{equation}
where ${\mathbf x}_\pm={\mathbf x}\pm\boldsymbol\epsilon/2$. Some straightforward algebra yields
\begin{equation}
C^{\,0}_{\alpha\alpha'}=-\frac{\hbar^2}{m}\nabla_x\cdot\nabla_\epsilon[\hat{\psi}^\dagger_\alpha({\mathbf x}_+)\hat{\psi}_{\alpha'}({\mathbf x}_-)]+[V_\alpha({\mathbf x}_+)-V_{\alpha'}({\mathbf x}_-)+E_\alpha-E_{\alpha'}]\,\hat{\psi}^\dagger_\alpha({\mathbf x}_+)\hat{\psi}_{\alpha'}({\mathbf x}_-)\,.
\label{eq:5.3S}
\end{equation}
Here, for spin-dependent harmonic potentials with spring constants  $k_{i\alpha}$,
\begin{equation}
V_\alpha({\mathbf x}_\pm)=
V_\alpha({\mathbf x})\mp\frac{\boldsymbol\epsilon}{2}\cdot F_\alpha({\mathbf x})+\frac{1}{8}\sum_i k_{i\alpha}\epsilon_i^2\,,
\label{eq:8.1S}
\end{equation}
where $F_\alpha({\mathbf x})=-\nabla_x V_\alpha({\mathbf x})$ is the spin dependent force. The interaction term is 
\begin{eqnarray}
C^{\,int}_{\alpha\alpha'}({\mathbf x}_+,{\mathbf x}_-)&=&-g_{12}\left\{\delta_{\alpha'\downarrow}\bar{n}_\uparrow({\mathbf x}_-)\hat{\psi}^\dagger_\alpha({\mathbf x}_+)\hat{\psi}_\downarrow({\mathbf x}_-)+\delta_{\alpha'\uparrow}\bar{n}_\downarrow({\mathbf x}_-)\hat{\psi}^\dagger_\alpha({\mathbf x}_+)\hat{\psi}_\uparrow({\mathbf x}_-)\right.\nonumber \\
& &\hspace*{0.5in}\left. -\delta_{\alpha'\downarrow}\bar{\rho}_{\uparrow\downarrow}({\mathbf x}_-)\hat{\psi}^\dagger_\alpha({\mathbf x}_+)\hat{\psi}_\uparrow({\mathbf x}_-)-\delta_{\alpha'\uparrow}\bar{\rho}_{\downarrow\uparrow}({\mathbf x}_-)\hat{\psi}^\dagger_\uparrow({\mathbf x}_+)\hat{\psi}_\downarrow({\mathbf x}_-)\right\}\nonumber\\
& &\hspace*{0.5in}+g_{12}\left\{\delta_{\alpha\downarrow}\bar{n}_\uparrow({\mathbf x}_+)\hat{\psi}^\dagger_{\downarrow}({\mathbf x}_+)\hat{\psi}_{\alpha'}({\mathbf x}_-)+\delta_{\alpha\uparrow}\bar{n}_\downarrow({\mathbf x}_+)\hat{\psi}^\dagger_\uparrow({\mathbf x}_+)\hat{\psi}_{\alpha'}({\mathbf x}_-)\right.\nonumber \\
& &\hspace*{0.5in}\left. -\delta_{\alpha\downarrow}\bar{\rho}^*_{\uparrow\downarrow}({\mathbf x}_+)\hat{\psi}^\dagger_\uparrow({\mathbf x}_+)\hat{\psi}_{\alpha'}({\mathbf x}_-)-\delta_{\alpha\uparrow}\bar{\rho}^*_{\downarrow\uparrow}({\mathbf x}_+)\hat{\psi}^\dagger_\downarrow({\mathbf x}_+)\hat{\psi}_{\alpha'}({\mathbf x}_-)\right\}.
\label{eq:5.5S}
\end{eqnarray}

Using these results, eq.~\ref{eq:4.2S} yields 
\begin{eqnarray}
& &\left[\partial_t+\frac{\mathbf p}{m}\cdot\nabla_x +\frac{{\mathbf F}_\alpha({\mathbf x})+{\mathbf F}_{\alpha'}({\mathbf x})}{2}\cdot\nabla_p\right]\,\hat{W}_{\alpha ,\alpha'}({\mathbf x},{\mathbf p})=i\,\Omega_{\alpha \alpha'}\,\hat{W}_{\alpha ,\alpha'}({\mathbf x},{\mathbf p})\nonumber\\
& &\hspace{0.25in}-\frac{\hbar}{8}\sum_i(k_{i\alpha}-k_{i\alpha'})\,\partial^2_{p_i}\hat{W}_{\alpha ,\alpha'}({\mathbf x},{\mathbf p})
+\frac{i}{\hbar}\,\int\frac{d^3 \boldsymbol\epsilon}{(2\pi\hbar)^3}\,e^{\frac{i\boldsymbol\epsilon\cdot{\mathbf p}}{\hbar}}\,C^{\,int}_{\alpha\alpha'}({\mathbf x}_+,{\mathbf x}_-)\,,
\label{eq:9.1S}
\end{eqnarray}
where the time argument is suppressed in the Heisenberg operators and 
\begin{equation}
\Omega_{\alpha\alpha'}=\omega_{\alpha\alpha'}+\frac{1}{\hbar}[V_\alpha({\mathbf x})-V_{\alpha'}({\mathbf x})]\,.
\label{eq:9.6S}
\end{equation}

The evolution equation for the thermal averaged density matrix $\bar{\rho}_{\alpha\alpha'}(\mathbf{x},t)$ is found by integrating eq.~\ref{eq:9.1S} over ${\mathbf p}$ using
\begin{equation}
\int\frac{d^3 {\mathbf p}}{(2\pi\hbar)^3}\,e^{\frac{i\boldsymbol\epsilon\cdot{\mathbf p}}{\hbar}}=\delta(\boldsymbol\epsilon)\,.
\label{eq:9.2S}
\end{equation}
With $\int d^3\boldsymbol\epsilon\delta(\boldsymbol\epsilon)\,C^{\,int}_{\alpha\alpha'}({\mathbf x}_+{\mathbf x}_-)=C^{\,int}_{\alpha\alpha'}({\mathbf x},{\mathbf x})$, we can see from eq.~\ref{eq:5.5S} that the thermal average $\langle C^{\,int}_{\alpha\alpha'}({\mathbf x},{\mathbf x})\rangle=0$. Thermal averaging  yields
 \begin{equation}
 \partial_t\bar{\rho}_{\alpha\alpha'}(\mathbf{x},t)+\nabla_x\cdot{\mathbf J}_{\alpha\alpha'}(\mathbf{x},t)=i\,\Omega_{\alpha\alpha'}\,\bar{\rho}_{\alpha\alpha'}(\mathbf{x},t)\,,
\label{eq:9.5S}
\end{equation}
where ${\mathbf J}_{\alpha\alpha'}$ is the  current density for each component of the density matrix.

To find the evolution equation for the current densities, we multiply eq.~\ref{eq:9.1S} by ${\mathbf p}/m$ and integrate over ${\mathbf p}$. We define  $\int d^3{\mathbf p}\,p_i\,p_j\, W_{\alpha ,\alpha'}({\mathbf x},{\mathbf p},t)=\langle p_i\,p_j \rangle_{\alpha\alpha'}\,\bar{\rho}_{\alpha\alpha'}(\mathbf{x},t)$.  Assuming that the momentum distribution remains close to thermal equilibrium and is nearly isotropic~\cite{TimeDeppsq},  $\langle p_i\,p_j\rangle_{\alpha\alpha'}\simeq\delta_{ij}\langle{\mathbf p}^2\rangle/3$ is independent of $\alpha\,,\alpha'$, $\mathbf{x}$, and $t$. The force terms are then evaluated using integration by parts, yielding
\begin{eqnarray}
& &\partial_t{\mathbf J}_{\alpha\alpha'}({\mathbf x},t)+\frac{\langle{\mathbf p}^2\rangle}{3m^2}\,\nabla_x \bar{\rho}_{\alpha ,\alpha'}({\mathbf x},t) -\frac{{\mathbf F}_\alpha({\mathbf x})+{\mathbf F}_{\alpha'}({\mathbf x})}{2m}\,\bar{\rho}_{\alpha ,\alpha'}({\mathbf x},t)=\nonumber\\
& &\hspace{0.75in}=i\,\Omega_{\alpha\alpha'}\,{\mathbf J}_{\alpha\alpha'}({\mathbf x},t)+\dot{\mathbf J}^{int}_{\alpha\alpha'}\,.
\label{eq:11.4S}
\end{eqnarray}
The interaction term is 
\begin{equation}
\dot{\mathbf J}^{int}_{\alpha\alpha'}=\int d^3{\mathbf p}\,\frac{\mathbf p}{m}\,\frac{i}{\hbar}\int\frac{d^3 \boldsymbol\epsilon}{(2\pi\hbar)^3}\,e^{i\frac{\boldsymbol\epsilon\cdot{\mathbf p}}{\hbar}}\,\langle C^{\,int}_{\alpha\alpha'}({\mathbf x}_+,{\mathbf x}_-)\rangle\,.
\label{eq:11.5S}
\end{equation}

We evaluate eq.~\ref{eq:11.5S} in terms of thermal averaged Wigner operators, $\langle\hat{W}_{\alpha ,\alpha'}({\mathbf x},{\mathbf p})\rangle$ using a method similar to that of ref.~\cite{LewensteinDynLongRange}.  In eq.~\ref{eq:5.5S}, we note that each term of $C^{\,int}_{\alpha\alpha'}({\mathbf x}_+,{\mathbf x}_-)$ takes the form
$\bar{f}({\mathbf x}_\pm)\hat{\psi}^\dagger_\alpha({\mathbf x}_+)\hat{\psi}_{\alpha'}({\mathbf x}_-)$. We can write
\begin{equation}
\bar{f}({\mathbf x}_\pm)=\sum_{l=0}^\infty\frac{1}{l!}\left(\pm\frac{\boldsymbol\epsilon}{2}\cdot\nabla_y\right)^l\,\bar{f}({\mathbf y})|_{{\mathbf y}\rightarrow {\mathbf x}}
\label{eq:11.6S}
\end{equation}
and
\begin{equation}
e^{i\frac{\boldsymbol\epsilon\cdot{\mathbf p}}{\hbar}}\left(\pm\frac{\boldsymbol\epsilon}{2}\cdot\nabla_y\right)^l\,\bar{f}({\mathbf y})=\left(\pm\frac{\hbar}{i}\frac{\nabla_p\cdot\nabla_y}{2}\right)^l\,e^{i\frac{\boldsymbol\epsilon\cdot{\mathbf p}}{\hbar}}\bar{f}({\mathbf y})\,.
\label{eq:12.1S}
\end{equation}
Then, eq.~\ref{eq:11.5S} becomes
\begin{equation}
\dot{\mathbf J}^{int}_{\alpha\alpha'}=\sum_l\frac{1}{l!}\int d^3{\mathbf p}\,\frac{\mathbf p}{m}\,\frac{i}{\hbar}\left(\pm\frac{\hbar}{i}\frac{\nabla_p\cdot\nabla_y}{2}\right)^l\,\int\frac{d^3 \boldsymbol\epsilon}{(2\pi\hbar)^3}\,e^{i\frac{\boldsymbol\epsilon\cdot{\mathbf p}}{\hbar}}\, \bar{C}^{\,int}_{\alpha\alpha'}({\mathbf y},{\mathbf x}_+,{\mathbf x}_-)|_{{\mathbf y}\rightarrow {\mathbf x}}\,,
\label{eq:12.2S}
\end{equation}
where the signs $\pm$ and the ${\mathbf y}$ argument in $\bar{C}^{\,int}$ correspond to the argument ${\mathbf x}_\pm$ of the first factor for each term in eq.~\ref{eq:5.5S}.

The $l=0$ terms in eq.~\ref{eq:12.2S} are proportional to ${\mathbf J}_{\alpha\alpha'}$. For the $l=1$ terms, integration by parts removes the ${\mathbf p}$ factor, yielding contributions proportional to gradients of the density matrix. For $l\geq 2$, the integrals vanish. 

We find the evolution equations for $\bar{\rho}_{\uparrow\downarrow}$ and $\bar{\rho}_{\uparrow\uparrow}$ from eq.~\ref{eq:9.5S}. Using eqs.~\ref{eq:11.4S}~and~\ref{eq:12.2S}, it is straightforward to find the evolution equations for ${\mathbf J}_{\uparrow\downarrow}$ and ${\mathbf J}_{\uparrow\uparrow}$. By interchanging $\uparrow$ and $\downarrow$, these results determine the evolution equations for all components of the spin density ${\mathbf s}$ using
\begin{eqnarray}
s_x&=&\frac{\bar{\rho}_{\uparrow\downarrow}+\bar{\rho}_{\downarrow\uparrow}}{2}\nonumber\\
s_y&=&\frac{\bar{\rho}_{\uparrow\downarrow}-\bar{\rho}_{\downarrow\uparrow}}{2i}\nonumber\\
s_z&=&\frac{\bar{\rho}_{\uparrow\uparrow}-\bar{\rho}_{\downarrow\downarrow}}{2}
=\frac{\bar{n}_\uparrow-\bar{n}_\downarrow}{2}\,.
\label{eq:10.1S}
\end{eqnarray}
and similarly for the components of the spin current tensor ${\mathbf J}_{x}$, ${\mathbf J}_{y}$, and ${\mathbf J}_{z}$.

We write the evolution equations in a frame rotating about the z-axis at the hyperfine resonance frequency, $\omega_{\uparrow\downarrow}$. In this frame,  eq.~\ref{eq:9.6S} requires  $\Omega_{\uparrow\downarrow}\rightarrow\Omega_z=-\Omega_{\downarrow\uparrow}$ with $\Omega_z\equiv\frac{1}{\hbar}[V_\uparrow({\mathbf x})-V_{\downarrow}({\mathbf x})]$ determined from eq.~\ref{eq:potentialsS} and $\Omega_{\uparrow\uparrow}=\Omega_{\downarrow\downarrow}=0$. Then, the spin density components $s_i=(s_x,s_y,s_z)$  obey
\begin{equation}
\dot{s}_i({\mathbf x},t)+\nabla\cdot{\mathbf J}_i({\mathbf x},t)=(\boldsymbol\Omega\times{\mathbf s})_i\,,
\label{eq:10.7S}
\end{equation}
where the spatially varying rotation rate is
\begin{equation}
{\mathbf \Omega}=\hat{\mathbf z}\,\frac{m\,\omega_0\delta\omega_{\uparrow\downarrow}}{\hbar}\,(x-x_{os})^2\,.
\label{eq:OmegaS}
\end{equation}

Now we explicitly write the evolution equations for the current densities ${\mathbf J}_i$ in eq.~\ref{eq:10.7S}, which enables a clear discussion of the negligible contributions. The x-component of the current density tensor obeys
\begin{equation}
\dot{\mathbf J}_x+\frac{\langle{\mathbf p}^2\rangle}{3m^2}\,\nabla s_x-\frac{{\mathbf F}_\uparrow+{\mathbf F}_\downarrow}{2m}\,s_x=-\Omega_z{\mathbf J}_y+\dot{\mathbf J}^{int}_x\,,
\label{eq:13.3S}
\end{equation}
where
\begin{equation}
\dot{\mathbf J}^{int}_x=-\frac{2g_{12}}{\hbar}(s_y\,{\mathbf J}_z-s_z\,{\mathbf J}_y)+\frac{g_{12}}{2m}(\bar{n}\nabla s_x-s_x\nabla\bar{n})\,,
\label{eq:13.2S}
\end{equation}
with  the total density $\bar{n}=\bar{n}_\uparrow+\bar{n}_\downarrow$.

Similarly, the y-component of the current density tensor obeys
\begin{equation}
\dot{\mathbf J}_y+\frac{\langle{\mathbf p}^2\rangle}{3m^2}\nabla s_y-\frac{{\mathbf F}_\uparrow+{\mathbf F}_\downarrow}{2m}\,s_y=\Omega_z{\mathbf J}_x+\dot{\mathbf J}^{int}_y\,,
\label{eq:14.3S}
\end{equation}
where
\begin{equation}
\dot{\mathbf J}^{int}_y=-\frac{2g_{12}}{\hbar}(s_z\,{\mathbf J}_x-s_x\,{\mathbf J}_z)+\frac{g_{12}}{2m}(\bar{n}\nabla s_y-s_y\nabla\bar{n})\,.
\label{eq:14.2S}
\end{equation}

Finally, the z-component of the current density tensor obeys
\begin{equation}
\dot{\mathbf J}_z+\frac{\langle{\mathbf p}^2\rangle}{3m^2}\nabla s_z-\frac{{\mathbf F}_\uparrow+{\mathbf F}_\downarrow}{2m}\,s_z=\frac{{\mathbf F}_\uparrow-{\mathbf F}_\downarrow}{2m}\,\frac{\bar{n}}{2}+\dot{\mathbf J}^{int}_z\,,
\label{eq:16.4S}
\end{equation}
where
\begin{equation}
\dot{\mathbf J}^{int}_z=-\frac{2g_{12}}{\hbar}(s_x\,{\mathbf J}_y-s_y\,{\mathbf J}_x)+\frac{g_{12}}{2m}(\bar{n}\nabla s_z-s_z\nabla\bar{n})\,.
\label{eq:16.2S}
\end{equation}

The evolution equation for the total density $\bar{n}=\bar{n}_\uparrow+\bar{n}_\downarrow$ is needed to complete the evolution equations. With ${\mathbf J}_n\equiv{\mathbf J}_{\uparrow\uparrow}+{\mathbf J}_{\downarrow\downarrow}$, we obtain the continuity equation
\begin{equation}
\dot{\bar{n}}+\nabla\cdot{\mathbf J}_n=0\,.
\label{eq:17.7S}
\end{equation}
The current density obeys
\begin{equation}
\dot{\mathbf J}_n+\frac{\langle{\mathbf p}^2\rangle}{3m^2}\nabla\bar{n}-\frac{{\mathbf F}_\uparrow+{\mathbf F}_\downarrow}{2m}\,\bar{n}=\frac{{\mathbf F}_\uparrow-{\mathbf F}_\downarrow}{m}\,s_z+\dot{\mathbf J}_n^{int}\,,
\label{eq:17.6S}
\end{equation}
where
\begin{equation}
\dot{\mathbf J}_n^{int}=\frac{g_{12}}{m}\nabla\left({\mathbf s}^2-\frac{\bar{n}^2}{4}\right).
\label{eq:17.4S}
\end{equation}

Using eq.~\ref{eq:17.6S}, it is easy to show that the density profile $\bar{n}({\mathbf x})$ is time-independent, as observed in our experiments. First, the differences between the spin-up and spin-down forces is negligible for our experimental parameters, as noted in the main text. Further, just after the $\pi/2$ radiofrequency pulse is applied, $s_x=\bar{n}/2$ is the only nonvanishing spin density component, so that $\dot{\mathbf J}_n^{int}=0$ in eq.~\ref{eq:17.4S}. Hence, the right hand side of eq.~\ref{eq:17.6S} initially vanishes. 
In thermal equilibrium, we expect that the total density is determined by the balance of the pressure $p$ and trap forces, $\nabla p+{\mathbf F}\bar{n}=0$. With ${\mathbf F}= ({\mathbf F}_\uparrow+{\mathbf F}_\downarrow)/2$ the average force and the pressure $p=\bar{n} m\langle v_x^2\rangle$, the sum of the second and third terms on the left side of eq.~\ref{eq:17.6S} also vanish. Then both $\dot{\mathbf J}_n=0$ and ${\mathbf J}_n=0$ just after the $\pi/2$ pulse, so that $\dot{\bar{n}}=0$ for all $t$. In modeling the data, we find that spatial derivatives of the mean field potential are negligible, i.e., the second terms on the right hand side of eqs.~\ref{eq:13.2S},~\ref{eq:14.2S},~and~\ref{eq:16.2S} are negligible. Further, the  force difference term on the right hand side of eq.~\ref{eq:16.4S} is negligible. 

\subsubsection{One-Dimensional Approximation}

As our experiments are performed in a cigar-shaped trap oriented along the x-axis with $\omega_\perp\simeq 25\,\omega_0$, we simplify the description further by making a one-dimensional approximation, using transverse-averaged densities and currents that depend only on $x$, with an average force near the cloud center $F_x=-m\omega_0^2 x$.  We take $\nabla\rightarrow\partial_x$ in the evolution equations and define a spin current density vector ${\mathbf J}\equiv (J_x,J_y,J_z)$, where the vector denotes the direction of the spin carried by the current flowing along x. Assuming a near equilibrium momentum distribution as discussed above, we take $\langle{\mathbf p}^2/3\rangle=\langle p^2\rangle=m\,k_BT$, where $T$ is the effective temperature. Dropping spatial derivatives of the mean field potential and force difference terms as discussed above, we obtain the simplified 1D evolution equations, 
\begin{equation}
\dot{\mathbf s}+\partial_x\,{\mathbf J}={\mathbf \Omega}\times{\mathbf s},
\label{eq:1.1simpS}
\end{equation}
\begin{equation}
\dot{\mathbf J}+\frac{k_BT}{m}\partial_x\,{\mathbf s}+\omega_0^2 x\,{\mathbf s}={\mathbf \Omega}\times{\mathbf J}-\frac{8\pi \hbar a_s}{m}\,{\mathbf s}\times{\mathbf J}\,,
\label{eq:1.2simpS}
\end{equation}
where ${\mathbf \Omega}$ is given by eq.~\ref{eq:OmegaS}. 

To model the evolution after the initial $\pi/2$ radio-frequency pulse, we take the only non-vanishing initial condition for the spin density to be $s_x(x,0)=\frac{1}{2}\,n_0(x)$. For simplicity,  we fit the density profile with $n_0(x)=\bar{n}_0\,\exp (-x^2/\sigma_x^2)$, which defines the effective temperature $k_BT=m\omega_0^2\sigma_x^2/2=\langle p^2/m\rangle$. The one-dimensional approximation implies that the strength of the  ${\mathbf s}\times{\mathbf J}$ interaction term, which is proportional to the square of the density, should be averaged over the transverse dimensions. For this reason, we take $\bar{n}_0$ to be a transverse averaged peak 3D density. Numerical integration determines the spin density profiles $n_{\uparrow,\downarrow}(x,t)=\frac{1}{2}\,n_0(x)\pm s_z(x,t)$ and the corresponding centers of mass
\begin{equation}
\langle x\rangle_{\uparrow\downarrow} =\frac{1}{N_{\uparrow\downarrow}}\int dx\, x\, n_{\uparrow\downarrow}(x,t)=\pm\frac{1}{N_{\uparrow\downarrow}}\int dx\, x\, s_z(x,t)\,,
\label{eq:CMS}
\end{equation}
where
\begin{equation}
N_{\uparrow\downarrow}\equiv\int dx\,n_{\uparrow\downarrow}(x,t)=\frac{\sqrt{\pi}}{2}\,\bar{n}_0\,\sigma_x\,.
\label{eq:Nup}
\end{equation}
Eq.~\ref{eq:CMS} shows that $\langle x\rangle_\downarrow=-\langle x\rangle_\uparrow$ for a 50-50 mixture, as predicted previously~\cite{LewensteinDynLongRange} and observed in our experiments.

\subsubsection{Center of Mass Oscillation}
Using eqs.~\ref{eq:1.1simpS},~\ref{eq:1.2simpS}~and~\ref{eq:CMS}, it is easy to show that $\langle x\rangle_{\uparrow\downarrow}$ obeys a driven oscillator equation. The z-component of eq.~\ref{eq:1.1simpS} requires $\dot{s}_z+\partial_xJ_z=0$. Multiplying by $x$ and integrating by parts
we obtain $N_\uparrow\partial_t\langle x\rangle_\uparrow=\int dx\,J_z$ and $N_\uparrow\partial^2_t\langle x\rangle_\uparrow=\int dx\,\dot{J}_z$. Integrating the z-component of eq.~\ref{eq:1.2simpS} and using eq.~\ref{eq:CMS} then shows that the center of mass of the spin-up component obeys a driven harmonic oscillator equation with a spin-dependent driving force, which changes sign for the spin-down component,
\begin{equation}
\partial^2_t\langle x\rangle_\uparrow +\omega_0^2 \langle x\rangle_\uparrow=-\frac{8\pi \hbar a_s}{m N_\uparrow}\int\! dx\, ({\mathbf s}\times{\mathbf J})_z\,.
\label{eq:1.12OscS}
\end{equation}

\begin{figure}[htb]
\includegraphics[width=5.5in]{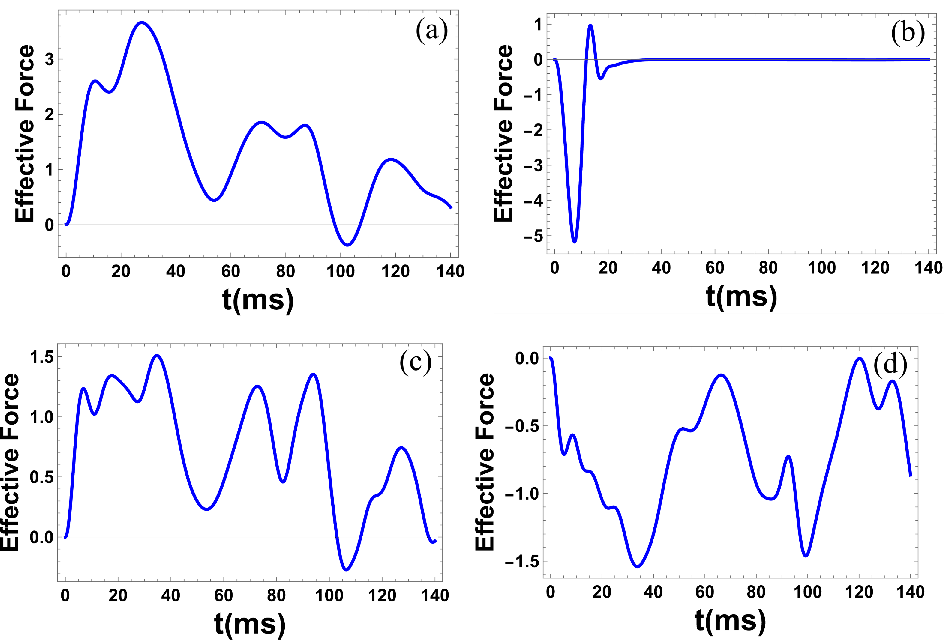}
    \caption{Effective force profiles, $\int\! dx\, ({\mathbf s}\times{\mathbf J})_z/\bar{n}_0^2$  $(10^{-3}cm^2/s)$, versus chirality and mean field strength. (a) Chirality parameter $x_{os} = 0.8\,\sigma_x$ with  $\Omega_0=52.6\,s^{-1}$ (eq.~\ref{eq:A5}) and mean field frequency $\Omega_{MF}=-334\,s^{-1}$ (eq.~\ref{eq:OmegaMF});  (b) Chirality parameter $x_{os} = -3.0\,\sigma_x$  with $\Omega_0=51.0\,s^{-1}$ and $\Omega_{MF}=-156\,s^{-1}$; (c) Chirality parameter $x_{os} = 1.15\,\sigma_x$  with $\Omega_0=53.8\,s^{-1}$ and $\Omega_{MF}=-586\,s^{-1}$; (d) Chirality parameter $x_{os} = -0.8\,\sigma_x$  with $\Omega_0=65.0\,s^{-1}$ and $\Omega_{MF}=-775\,s^{-1}$. Note that for $a_s<0$, the sign of the force on $\langle x\rangle_\uparrow$ is $+$ for $x_{os}>0$ (a,c), and $-$ for $x_{os}<0$ (b,d). Figs.~\ref{fig:ImpulsiveForce}(a-d) correspond to Figs. 3a, 3d, 4, and 5 of the main text.}
    \label{fig:ImpulsiveForce}
\end{figure}

To characterize the relative strength of the chirality and the scattering interactions, we rewrite the spatially varying rotation rate of eq.~\ref{eq:OmegaS} as
\begin{equation}
{\mathbf \Omega}=\hat{\mathbf z}\,\Omega_0\left(\frac{x-x_{os}}{\sigma_x}\right)^2\,,
\label{eq:A5}
\end{equation}
where the frequency $\Omega_0=m\omega_0\delta\omega_{\uparrow\downarrow}\sigma_x^2/\hbar$  and the offset $x_{os}$ determine the twist rotation rate~\cite{Deltaomega}. The  interaction strength is characterized by the mean field frequency,
\begin{equation}
\Omega_{MF}=\frac{4\pi\hbar a_s \bar{n}_0}{m}\,.
\label{eq:OmegaMF}
\end{equation}

We find by numerical integration of eqs.~\ref{eq:1.1simpS}~and~\ref{eq:1.2simpS} that the driving term $\int\!dx\,({\mathbf s}\times{\mathbf J})_z$ in eq.~\ref{eq:1.12OscS} generally has a complicated shape,  which modulates the amplitude of the $\langle x\rangle_\uparrow$ oscillation.  Fig.~\ref{fig:ImpulsiveForce}(a) shows this behavior for a chirality offset parameter $x_{os}\simeq 0.8\,\sigma_x$, at $B=511$ G, where $a_s=-53.1\,a_0$. Similar behavior is observed in Fig.~\ref{fig:ImpulsiveForce}(c) for $x_{os}=1.15\,\sigma_x$ at $B=500$ G, where $a_s=-84.6\,a_0$, and in Fig.~\ref{fig:ImpulsiveForce}(d) for $x_{os}=-0.8\,\sigma_x$ at $B=437$ G, where $a_s=-213.1\,a_0$.  For these parameters, see Table~\ref{Parameters} below, the mean field rotation rate $\Omega_{MF}$ is comparable to or larger than the spread $\simeq 2\,\Omega_0$ in the twist rotation rate $\Omega_z$ and larger than the trap frequency $2\pi\nu_x$. 
The effective force has a significant slowly decaying (``DC") contribution that shifts the $\uparrow\,,\downarrow$ centers of oscillation in opposite directions to produce an apparent bouncing behavior in $\langle x\rangle_{\uparrow\,,\downarrow}$. 

\begin{figure}[htb]
\hspace{-0.2in}\includegraphics[width=5.5in]{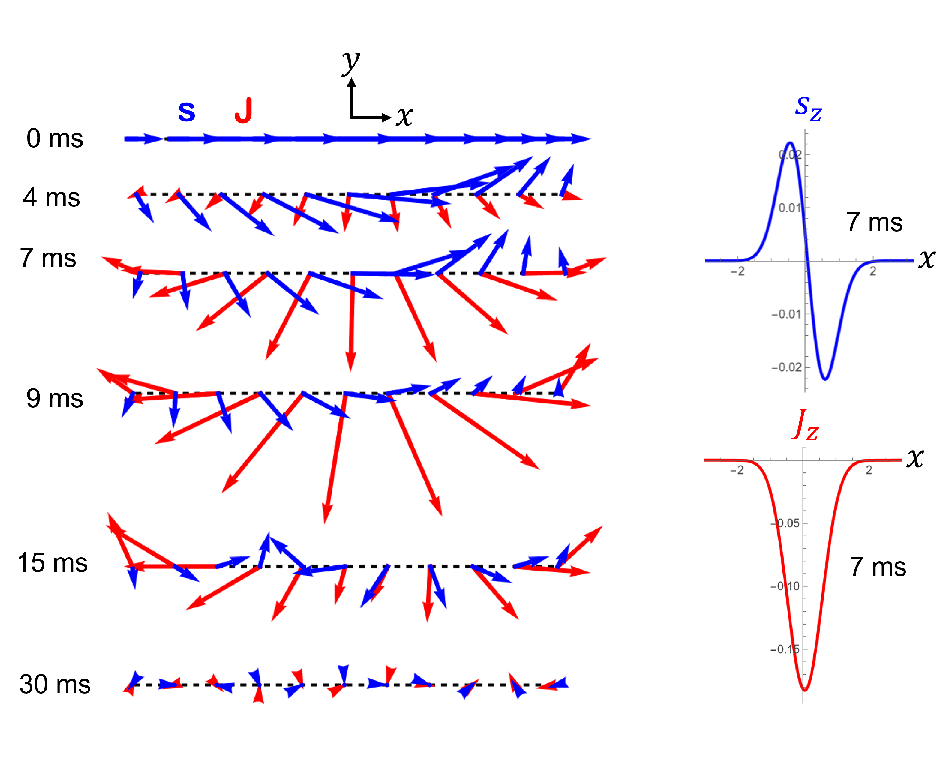}
     \caption{Spin density vector ${\mathbf s}$ (blue) and spin current density vector ${\mathbf J}$ (red) versus time for $x_{os}=-3.0\,\sigma_x$. The direction of ${\mathbf J}$ denotes the direction of the spin carried by the current, which flows along the $x$-axis. Components are given in a frame rotating about the $z$-axis at the mean frequency $\langle\Omega_z(x)\rangle$ relative to the hyperfine frequency.}
    \label{fig:VectorPlots}
\end{figure}
\begin{figure}[htb]
\hspace{-0.2in}\includegraphics[width=5.5in]{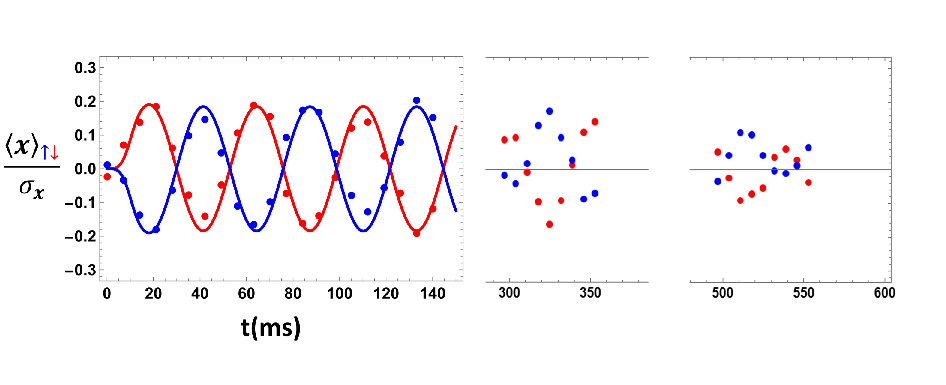}
    \caption{Oscillation of the center of mass for each spin state for $a_s=-53.1\,a_0$ and $\bar{n}_0=0.42/\mu m^3$  with $x_{os} = -3.0\,\sigma_x$, corresponding to the effective force of Fig.~\ref{fig:ImpulsiveForce}(b).  Model (solid curves). }
    \label{fig:511GM3OscHalfsec}
\end{figure}

\newpage
The opposite regime is shown in Fig.~\ref{fig:ImpulsiveForce}(b) for a large offset $x_{os}=-3\,\sigma_x$ with $a_s=-53.1\,a_0$. Here, the spatially varying rotation rate $\Omega_z$ in the trap is nearly linear in $x$ with a spread in frequency $\simeq 12\,\Omega_0$ across the cloud that is large compared to both the trap frequency $2\pi\nu_x$ and the mean field frequency $\Omega_{MF}$ (see Table~\ref{Parameters} below), dominating the evolution of the spin vectors and spin currents. As shown in Fig.~\ref{fig:VectorPlots}, after the $\pi/2$ pulse, the spin current density scatters from a transient spin spiral density in the $x-y$ plane,  which forms and vanishes quickly compared to the trap period. Initially, ${\mathbf J}=0$ and $\dot{\mathbf J}=0$ for the initial $x$-polarized spin density ${\mathbf s}(x,0)$.  For times $t$ short compared to the trap and mean field periods,   ${\mathbf s}(x,0)$ is rotated counterclockwise (for $x_{os}<0$) by an angle $\propto x$. The $x$-dependent angle produces a spin spiral density ${\mathbf s}(x,t)$  for which $\dot{\mathbf J}\neq 0$ and ${\mathbf s}(x,t)$ are both polarized in the $x-y$ plane and nominally perpendicular. For $a_s<0$, the s-wave scattering rate $\propto - a_s\,{\mathbf s}\times{\mathbf J}$  produces a spin current with a negative $z$-component $J_z$, after which the rapidly increasing spread in angles causes the $x$ and $y$ components of ${\mathbf s}$ and ${\mathbf J}$ to vanish. S-wave scattering is forbidden for the remaining $z$-polarized  spin current and spin density, which continue to oscillate freely. For this case, the right hand side of eq.~\ref{eq:1.12OscS} is an impulse just after $t=0$, with a negligible ``DC" component,  Fig.~\ref{fig:ImpulsiveForce}(b), leading to sinusoidal oscillation about $x=0$, Fig.~\ref{fig:511GM3OscHalfsec}.  We observe that the oscillation period is close to the trap period $1/\nu_x=45.8$ ms, and persists for times scales of  $0.5$ s with a slow decay, which we do not model.

\subsubsection{Model Fit Parameters}

In the fits using the 1D model, we initially estimate the radially averaged 3D density from the peak 3D density for the assumed gaussian spatial profile,  $n_0=N/(\pi^{3/2}\sigma_x\sigma_r^2)$, where $N$  is the total atom number, $\sigma_x$ is the axial 1/e width, and $\sigma_r$ is the transverse 1/e radius. We estimate $\sigma_r=\sigma_x\,\nu_x/\nu_r$, where $\nu_x$ is the axial oscillation frequency. With a transverse oscillation frequency of $\nu_r=600$ to $800$ Hz, dependent on the optical trap depth, we find that $n_0\simeq 1/\mu m^3$ in all cases. Then we choose $\bar{n}_0$ in the range from $n_0/2$ and $n_0$ to obtain the fits shown. The predictions are quite sensitive to the axial trap frequency, which is adjusted by a few percent to optimize the fit. In Table~\ref{Parameters}, we summarize the parameters used to obtain the predictions shown in the figures for the main text.  

\begin{table}[h]
\caption{Fit Parameters. Magnetic field $B$(G); $^6$Li s-wave scattering length $a_s=a_{12}(a_0)$;    Total atom number $N(10^3)$; Axial gaussian $1/e$ radius $\sigma_x(\mu m)$;  Trap axial frequency $v_x$(Hz); Fitted average peak density $\bar{n}_0(\mu m^{-3})$; Spin rotation frequency $\Omega_0(s^{-1})$;  Offset $x_{os}(\sigma_x)$; Mean field frequency $\Omega_{MF}(s^{-1})$. \label{Parameters}}
\vspace*{0.25in}
\begin{tabular}{|c|c|c|c|c|c|c|c|c|c|}
            \hline
          Figure &$B$ &$a_{12}$ & $N$ &$\sigma_x$ & $\nu_x$ &  $\bar{n}_0$ & $\Omega_0$ & $x_{os}$& $\Omega_{MF}$ \\
             \hline
            2a&511&-53.1   &76 & 198& 25.9& 0.90 & 52.6& 1.0 &-334\\
             2b&437&-213.1 & 65 & 195& 24.3& 0.52 & 65.0 & -0.9 &-775\\
            3a&511&-53.1  & 76  & 198 &25.9 & 0.90& 52.6 & 0.8 &-334\\
            3b&511&-53.1 &  76 & 198 &25.9 & 0.90 & 52.6 & 0.0 &-334\\
             3c&511&-53.1 &  76 & 198 &25.9 & 0.90 &52.6 &-0.8 &-334\\
             3d&511&-53.1&  48 & 195 &21.8 & 0.42 & 51.0 & -3.0 &-156\\
             4&500&-84.6 & 70 & 197 &22.4 & 0.99 & 53.8 & 1.15 &-586\\
             5&437&-213.1 &  65 & 195 &24.3 & 0.52 & 65.0 & -0.8 &-775\\          
              \hline 
\end{tabular}
\end{table}


\begin{thebibliography}{34}
\bibitem [{BW()}]{BW}%
  \BibitemOpen
  \href@noop {} {}\bibinfo {note} {A 0.5 ms $\pi/2$ pulse assures uniform
  excitation, as the Rabi frequency $\Omega_R\simeq 3\times 10^3\,s^{-1}$ is
  large compared to the spatially varying detuning of eq.~\ref{eq:Omegaz},
  where $\Omega_0=52.6\,s^{-1}$.}\BibitemShut {Stop}%
\bibitem [{Del()}]{Deltaomega}%
  \BibitemOpen
  \href@noop {} {}\bibinfo {note} {Here,
  $\omega^2_\uparrow-\omega^2_\downarrow=-\,\omega^2_{\rm mag}\partial_B
  E_{\uparrow\downarrow}(B)/\mu_B\simeq
  2\omega_0\delta\omega_{\uparrow\downarrow}$, with $\mu_B$ the Bohr magneton
  and $-E_{\uparrow\downarrow}(B)=E_2(B)-E_1(B)$ the energy difference for the
  two lowest $^6$Li hyperfine states. $\omega_{\rm mag}$ is the measured
  oscillation frequency for either spin state in the magnetic potential. See
  ref.~\cite{SaeedPRASpinECorrel}.}\BibitemShut {Stop}%
\bibitem [{Spi()}]{SpinSpiral}%
  \BibitemOpen
  \href@noop {} {}\bibinfo {note} {Strictly speaking, a spin-spiral is produced
  by a $z$-dependent spin rotation about the z-axis, where $x\rightarrow z$ in
  the 1D model, giving identical behavior.}\BibitemShut {Stop}%
\bibitem [{Rev()}]{Reversal}%
  \BibitemOpen
  \href@noop {} {}\bibinfo {note} {In the initial experiments, we observed
  qualitatively that reversing the sign of the scattering length by tuning $B$
  above the zero crossing field $\simeq 527.18$ G~\cite{SaeedPRASpinECorrel}
  interchanges the roles of the spin-up and spin-down profiles for the same
  $x_{os}$.}\BibitemShut {Stop}%
\bibitem [{Err()}]{ErrorBar}%
  \BibitemOpen
  \href@noop {} {}\bibinfo {note} {Each data point is an average of $4$ shots,
  taken in random time order. Error bars denote the standard deviation of the
  mean.}\BibitemShut {Stop}%
\bibitem [{Tim()}]{TimeDeppsq}%
  \BibitemOpen
  \href@noop {} {}\bibinfo {note} {Although it has been employed in previous
  treatments, this assumption is not obvious in the weakly interacting, nearly
  collisionless regime. For a noninteracting gas, $a_{12}=0$, one can obtain
  analytic Wigner phase space distributions. In a one dimensional harmonic
  trap, $W_{\alpha,\alpha'}(x,p,t)$ has a position and momentum dependent phase
  for $\alpha'\neq\alpha$, arising from the spatially varying rotation rate,
  $\phi=\int_0^t dt'\,\Omega_z[x(x_0,p_0,t')]$, where the initial conditions
  $x_0$ and $p_0$ are determined by $x$ and $p$ at time $t$. With the
  definition, $\langle
  p^2\rangle_{\alpha,\alpha'}\,\rho_{\alpha,\alpha'}(x,t)=\int dp\,
  p^2\,W_{\alpha,\alpha'}(x,p,t)$, one then finds that $\langle
  p^2\rangle_{\uparrow,\downarrow}$ oscillates in time (and can be
  x-dependent), altering the equations of motion for the $x$ and $y$ components
  of the current densities, and hence the effective force. For
  $\alpha'=\alpha$, there is no phase, so the $z$ component of the current is
  unaffected. We find that a time dependent $\langle
  p^2\rangle_{\uparrow,\downarrow}$ is inconsistent with the data, in contrast
  to the 1D thermal approximation $\langle p^2\rangle_{\alpha,\alpha'}=mk_BT$,
  which yields good agreement.}\BibitemShut {Stop}%
\end{thebibliography}
\end{document}